\documentclass[pra,aps,showpacs,showkeys,raggedbottom,superscriptaddress,tightenlines,amssymb,twocolumn,longbibliography]{revtex4-2}
\usepackage{graphicx}
\usepackage{amsmath}
\usepackage{amsfonts}
\usepackage{amssymb}
\usepackage{epsfig,libertine}
\usepackage[libertine]{newtxmath}
\usepackage{color}
\usepackage{dsfont, hyperref, xcolor}
\usepackage{comment}
\usepackage{enumerate}
\usepackage{babel}[english, italian]

\newcommand{\bra}[1]{\left\langle#1\right\vert}
\newcommand{\ket}[1]{\left\vert#1\right\rangle}

\hypersetup{colorlinks=true,urlcolor=red,citecolor=blue,filecolor=magenta}

\begin{document}

\title{Work extraction from coherently activated maps via quantum switch}

\author{Kyrylo Simonov}
\affiliation{Fakult\"{a}t f\"{u}r Mathematik, Universit\"{a}t Wien, Oskar-Morgenstern-Platz 1, 1090 Vienna, Austria}
\author{Gianluca Francica}
\affiliation{CNR-SPIN, Via Giovanni Paolo II 27, 84084 Fisciano (SA), Italy}
\author{Giacomo Guarnieri}
\affiliation{School of Physics, Trinity College Dublin, College Green, Dublin 2, Ireland}
\affiliation{Dahlem Center for Complex Quantum Systems, Freie Universit\"{a}t Berlin, Arnimallee 14, 14195 Berlin, Germany}
\author{Mauro Paternostro}
\affiliation{Centre for Theoretical Atomic, Molecular and Optical Physics, School of Mathematics and Physics, Queen's University Belfast, Belfast BT7 1NN, United Kingdom}

\date{\today}
\begin{abstract}
We characterize the impact that the application of two maps in a quantum-controlled order has on the process of work extraction via unitary cycles and its optimization. The control is based on the quantum switch model that applies maps in an order not necessarily compatible with the underlying causal structure and, in principle, can be implemented experimentally. First, we show that the activation of quantum maps through the quantum switch model always entails a non-negative gain in ergotropy compared to their consecutive application. We also establish a condition that the maps should fulfill in order to achieve a non-zero ergotropic gain. We then perform a thorough analysis of maps applied to a two-level system and provide general conditions for achieving a positive gain on the incoherent part of ergotropy. Our results are illustrated with several examples and applied to qubit and $d$-dimensional quantum systems. In particular, we demonstrate that a non-zero work can be extracted from a system thermalized by two coherently controlled reservoirs.
\end{abstract}

\maketitle

\section{Introduction}

One of central questions in thermodynamics is the maximum amount of work that can be extracted from a given medium. The extractable work is constrained by the second law, which establishes the {\it temporal direction} of any physical process the work medium undergoes. Interestingly, quantum mechanics allows in principle for superpositions of processes: for instance, occurrence of processes can be put under coherent control by a quantum system. A particular example is the {\it quantum switch model (QSM)}, which comprises a qubit used to control the order of quantum operations (maps)~\cite{Chiribella2013}. If two different time orderings of the same process are applied depending on which pure state the control qubit is, a coherent superposition will prepare a non-separable combination of such orderings. QSM can thus be seen as a realization of a process that is not compatible with a specific underlying causal structure~\cite{Oreshkov2012, Brukner2014, Araujo2015}. Due to its experimental appeal, the QSM has attracted high interest in the community and was used to demonstrate experimentally the advantages for communication~\cite{Rubino2017, Goswami2018, Goswami2020, Guo2020, GoswamiReview2020, Rubino2021}, computing~\cite{Procopio2015, GoswamiReview2020}, and thermodynamic tasks~\cite{Nie2020, Cao2021, Felce2021} if a system undergoes action of maps applied in a quantum-controlled order.

While emphasizing the disputed role of the causal inseparability as a resource~\cite{Guerin2019, Kristjansson2020, Chiribella2021, Rubino2021, Bhattacharya2021} is not a goal of this work, we intend to study how the coherent activation of the maps in the QSM allows to optimize certain thermodynamic tasks. Very recent work has addressed the implications that the QSM might have for the performance of cooling cycles~\cite{Felce2020, Goldberg2021, Nie2022} and work-extraction games~\cite{Guha2020}. In particular, when dealing with work extraction and heat dissipation resulting from quantum processes, attention should be paid to the role played by quantum coherence present or created in the state of the work medium~\cite{Francica2019, Santos2019}, which are crucial. On the other hand, the finite dimensions of the system and quantum effects put extra constraints on the effectivity of a device: in particular, the resulting maximal work that can be extracted from a medium does not necessarily coincide with its ``usual'' thermodynamic bound established by free energy. In this paper, hence, we make use of the concept of \textit{ergotropy}~\cite{Allahverdyan2014} that takes into account such constraints to understand the advantages of quantum control for work extraction and explore the interplay between quantum coherence in the work medium and quantum-controlled activation of the maps. Technically, ergotropy quantifies the maximal work that can be extracted from a quantum system through a cyclic unitary transformation of the reference Hamiltonian parameters. Our choice of figure of merit is due to recently unveiled links between ergotropy and quantum coherence~\cite{Cakmak2020, Francica2020, Gherardini2020, Diaz2020}, which we combine here with the use of quantum-controlled ordering of maps. Our goal is to show that the latter embodies a resource for ergotropic games.

Even in the simplest case when the QSM superposes the orders of application of two commutative maps, we find an enhancement of ergotropy compared with the corresponding probabilistic mixture realizing a well-defined order of them. Interestingly, the amount of quantum coherence initially present in the system plays an important role: for certain maps and system states, the QSM allows to enhance the ergotropy only by consuming coherence. Our findings pave the way to the assessment of finite-time (dis)charging protocols of quantum batteries~\cite{Binder2015}, where quantum operations able to seed quantum coherences in the state of a multi-particle medium appear to be advantageous, and which might benefit of the explicit use of processes occurring in a quantum-controlled order.

The remainder of this paper is organized as follows. In Sec.~\ref{Ergotropy}, we introduce the notion of ergotropy and identify the conditions for the system when it is not possible to extract work from it. In Sec.~\ref{QSM_Setup}, we introduce the setup which consists of the work medium previously undergone a combination of maps. In Sec.~\ref{QSM_DaemErgotropy}, we discuss how ergotropy is changed for quantum-controlled occurrence of maps within QSM. In Sec.~\ref{sec:GainConditions}, we provide the conditions for a non-zero gain in ergotropy under the quantum control of maps realized by the QSM. In Sec.~\ref{QubitErgotropy}, we consider a two-level quantum system sent through two commutative maps and study the amount of extractable work resulting from quantum-controlled and classical-controlled maps, respectively. In Sec.~\ref{Identmaps}, we particularize our findings and address the gain in ergotropy for a quantum-controlled occurrence of two identical completely depolarizing maps. In Sec.~\ref{identMapsTherm}, we generalize these findings to thermalizing maps and demonstrate the activation of the system's state under quantum-controlled occurrence of them even for a completely passive initial state. In Sec.~\ref{QubitGain}, we consider the example of a two-level quantum system sent through the commutative amplitude damping and phase flip channel and study the enhancement of ergotropic work, assessing the role that coherence plays. Finally, in Sec.~\ref{Conclusions}, we draw the conclusions and comment on potential further applications of our framework.

\section{Extractable work from quantum switch}\label{ErgotropyQSM}

\subsection{Preliminaries: Extractable work from a quantum system}\label{Ergotropy}

We consider the work medium as a quantum system $S$ being in a state $\hat{\rho}_S = \sum_k r_k |r_k\rangle\langle r_k|$ with the eigenstates sorted in decreasing order as $r_k \geq r_{k+1}$. The work is extracted from $S$ via a cyclic process (de)coupling it to/from external sources, whereas before and after work extraction the system $S$ is regarded isolated. Hence, dynamics of $S$ in a fixed time interval $t\in[0,\tau]$ is governed by the time-dependent Hamiltonian $\hat{H}_S(t)$ such that $\hat{H}_S(0) = \hat{H}_S(\tau) \equiv \hat{H}_S$, where the Hamiltonian $\hat{H}_S = \sum_k \epsilon_k |\epsilon_k\rangle\langle\epsilon_k|$ with the energies being sorted in increasing order as $\epsilon_k \leq \epsilon_{k+1}$ describes $S$ in a thermal isolation. In this scenario, the resulting evolution of $S$ appears to be unitary, and the maximal work that can be extracted from $S$ via such cyclical processes is given by the {\it ergotropy}~\cite{Allahverdyan2014}
\begin{equation}\label{eq:ergotropy}
    W(\hat{\rho}_S) = \max_{\hat{U} \in \mathcal{U}_\Box} \operatorname{Tr}\Bigl[\hat{H}_S (\hat{\rho}_S - \hat{U}\hat{\rho}_S\hat{U}^\dagger)\Bigr],
\end{equation}
where $\mathcal{U}_\Box$ denotes the set of unitary transformations generated by the time-dependent Hamiltonian under consideration. The optimal unitary cycle $\hat{U}_e \in \mathcal{U}_\Box$, which maximizes the extracted work, transforms $\hat{\rho}_S = \sum_{k,k'} \rho_{kk'} |\epsilon_k\rangle\langle \epsilon_k'|$ into a \textit{passive} state $\hat{U}_e\hat{\rho}_S\hat{U}^\dagger_e = \sum_k r_k |\epsilon_k\rangle\langle \epsilon_k|$, so that no more work can be extracted from $S$ cyclically. In this way, $W(\hat{\rho}_S) = 0$ if and only if the state of the work medium is passive, i.e.,
\begin{enumerate}[(a)]
    \item $\hat{\rho}_S$ and $\hat{H}_S$ commute,
    \item the populations $\rho_{kk} = \langle \epsilon_k | \hat{\rho}_S | \epsilon_k \rangle$ are sorted in decreasing order.
\end{enumerate}
Importantly, when the condition (a) is satisfied, the condition (b) means that the more energetic is the eigenstate, the lower is the probability to find $S$ in it.

\subsection{Setup}\label{QSM_Setup}

Extending previous approach we consider the situation when the work is extracted from a medium $S$ previously influenced by the environment, i.e., undergone action of certain completely positive and trace-preserving (CPTP) maps. The physics behind these CPTP maps can vary: for instance, work is intended to be extracted from $S$ that has been transmitted via noisy channels to the laboratory~\cite{Ebler2018, Cacciapuoti2019, Caleffi2020, Koudia2021}, undergone a thermalization process in the presence of a thermal bath~\cite{Guha2020}, or been charged as a quantum battery~\cite{Chen2021}.

\begin{figure}[t!]
    \includegraphics[width=\columnwidth]{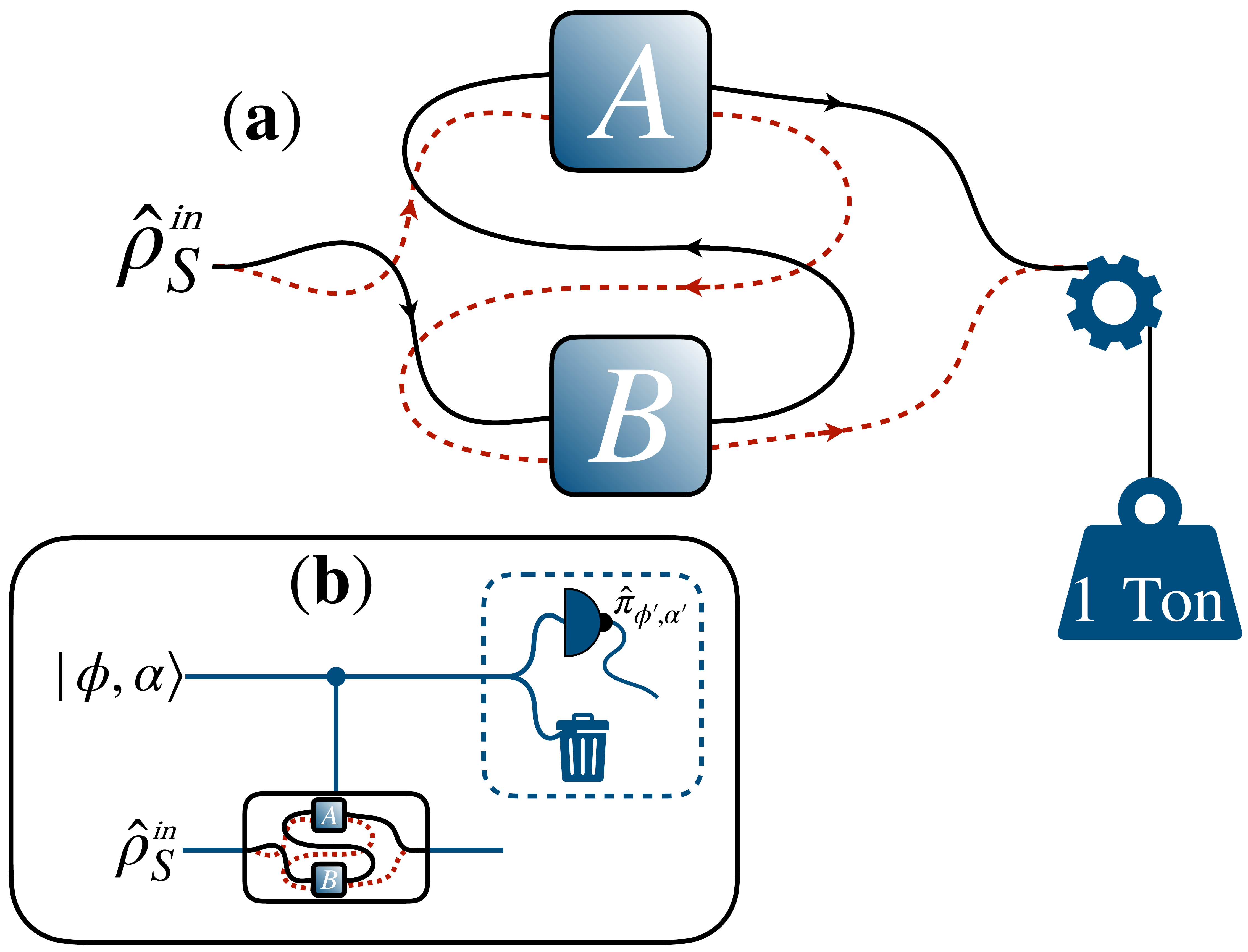}
    \caption{{\bf (a)} Scheme of the principle of the problem under scrutiny. A system $S$ is initially prepared in a state $\hat\rho^{in}_S$ and undergoes a certain composition of two CPTP maps, $\mathcal{A}$ and $\mathcal{B}$. The maximum work that can be extracted from the system is then evaluated. The red dashed line shows the configuration where the application of $\mathcal{A}$ to $S$ precedes that of $\mathcal{B}$, while the solid black line represents the opposite ordering of the maps. In the QSM, we consider a superposition of the two distinguishable configurations and investigate the effects that coherence plays in such an indefinite causal ordering. This is achieved as illustrated in panel {\bf (b)}, where the control qubit $Q$ prepared in state $\rho_Q$ (assumed to be pure in what follows, $\rho_Q \equiv \ket{\phi,\alpha}\bra{\phi,\alpha}$) induces a specific map-ordering conditional on its logical state [cf. main text]. The preparation of $Q$ in its logical $\ket{0}$ ($\ket{1}$) state induces the configuration corresponding to the solid black (dashed red) line. The resulting two-body evolution can thus be interpreted as a controlled gate. The state of $Q$ is then either discarded or projected onto state $\ket{\phi',\alpha'}$.}
    \label{schema}
\end{figure}

Remaining as general as possible we assume nevertheless that the action of the environment can be separated into two CPTP maps $\mathcal{A}[\hat{\rho}_S^{in}]$ and $\mathcal{B}[\hat{\rho}_S^{in}]$ that act on the initial state $\hat{\rho}_S^{in}$ of $S$ in some order that we are able to control, and the ergotropic work is extracted from the output state $\hat{\rho}_S \equiv \hat{\rho}_S^{out}$ of the overall combination of these maps (see Fig.~\ref{schema}(a)). If we imply well-defined causal relations between $\mathcal{A}$ and $\mathcal{B}$ then each map can act on the work medium only once. The maps can be placed thus in a certain consecutive order, namely, $(\mathcal{A} \circ \mathcal{B})[\hat\rho_S^{in}]$ or $(\mathcal{B} \circ \mathcal{A})[\hat\rho_S^{in}]$, or we can choose between them randomly, for example, by tossing a coin. In this case, we control the maps classically, and the ergotropic work is extracted from the state
\begin{equation}
\label{ClassicalControl}
\begin{aligned}\hat{\rho}^{out, class}_S &= \phi (\mathcal{A} \circ \mathcal{B})[\hat{\rho}^{in}_S] + (1-\phi) (\mathcal{B} \circ \mathcal{A})[\hat{\rho}^{in}_S],
\end{aligned}
\end{equation}
where $\phi\in[0,1]$ defines the probability of placing $\mathcal{A}$ and $\mathcal{B}$ in one or another causal order.

Quantum mechanics allows to go beyond this scenario and relax the implied causal relations between the maps by making their application order subject to superposition. This can be achieved with a coherent control of the maps' application by another quantum system $Q$ that assists the work medium $S$. Such control is realized within the QSM~\cite{Chiribella2013, Wechs2021} which picks a qubit in a state $\hat{\rho}_Q$ as the control system $Q$ and entangles it with the application orders of the maps. Our figure of merit is the extractable work gained from the placement of the CPTP maps $\mathcal{A}$ and $\mathcal{B}$ into the QSM from foundational and not only practical point of view. Hence, we consider the QSM as a black box (with an input and output for $S$) without referring to its particular implementation. Nevertheless, it is useful to sketch the possible physical scenarios that can be exploited for this setup. Within standard physics, when the CPTP maps are regarded as noisy channels used to transmit $S$ to the laboratory, interferometric setups allow to simulate the QSM. Therein, $S$ and $Q$ are carried by a photon and encoded on different degrees of freedom of it~\cite{Procopio2015, Rubino2017, Goswami2018, Guo2020, Rubino2021}. More exotic scenarios beyond standard physics could implement the QSM with a fundamental distinction between $S$ and $Q$. Such scenarios include the use of closed time-like curves (CTCs)~\cite{Chiribella2013} or superposed gravitational fields~\cite{Zych2019, Paunkovic2020}. Therein, $Q$ is determined independently of the sender of $S$ and does not involve any channel to transmit it. It is therefore reasonable to consider $\hat{\rho}_Q$ fixed in order to take into account the whole spectrum of potential implementations of the QSM and study its benefits independently of its practical realization.

Once applied, the QSM induces a conditional evolution on the system $S$ which depends on the initial state $\hat{\rho}_Q$ of the control qubit $Q$. Specifically, the resulting evolution of the joint setup $SQ$ is induced by a higher-order process represented as a supermap that sends $\mathcal{A}$ and $\mathcal{B}$ to another CPTP map $\mathbb{M}_{\hat{\rho}_Q}(\mathcal{A}, \mathcal{B})$,
\begin{equation}
\label{QSmap}
    \hat{\rho}_{SQ}^{out} = \mathbb{M}_{\hat{\rho}_Q}(\mathcal{A}, \mathcal{B})[\hat{\rho}_S^{in}]\color{black} = \sum_{ij} \hat{K}_{ij} (\hat{\rho}_S^{in} \otimes \hat{\rho}_Q) \hat{K}^\dagger_{ij},
\end{equation}
whose Kraus decomposition is given by the operators $\hat{K}_{ij} = \hat{A}_i \hat{B}_j \otimes |0\rangle \langle 0| + \hat{B}_j \hat{A}_i \otimes |1\rangle \langle 1|$, where the sets of operators $\{\hat{A}_i\}$ and $\{\hat{B}_j\}$ constitute the Kraus decompositions of $\mathcal{A}$ and $\mathcal{B}$, respectively. This physically implies that the control qubit $Q$ prepared in a computational basis state places the maps into a specifically ordered composition, namely $(\mathcal{A} \circ \mathcal{B})[\hat\rho_S^{in}]$ for $\hat{\rho}_Q = |0\rangle\langle 0|$ and $(\mathcal{B} \circ \mathcal{A})[\hat\rho_S^{in}]$ for $\hat{\rho}_Q = |1\rangle\langle 1|$, that is applied thereafter to the work medium $S$. As a consequence of that, any incoherent state of the control qubit $Q$ (with respect to $\hat{\sigma}_z$ basis), i.e., $\hat{\rho}_Q = \mathrm{diag}(\phi,1-\phi)$, realizes a convex combination of the composite maps in agreement with~(\ref{ClassicalControl}),
\begin{align}\label{SepChan}
    \hat{\rho}_{SQ}^{out, i} &= \phi\Bigl((\mathcal{A} \circ \mathcal{B})[\hat{\rho}_S^{in}] \otimes  |0\rangle \langle 0|\Bigr) \notag\\
    &\qquad + (1 - \phi)\Bigl((\mathcal{B} \circ \mathcal{A})[\hat{\rho}_S^{in}] \otimes  |1\rangle \langle 1|\Bigr).
\end{align}
In this particular case, the QSM behaves as the so-called ``classical switch'' that chooses between two possible application order of the maps with a certain probability defined by $\phi$~\cite{Wechs2021} and does not violate causality. On the other hand, any coherence in $Q$ inevitably results in a combination of the maps, not of the form of Eq.~\eqref{SepChan} and incompatible with any well-defined causal structure between them. It is now very natural to ask whether such a non-consecutive application of the maps due to a coherent superposition of the control qubit leads to an enhancement or a depletion in terms of the maximum amount of extractable work compared with the classical-controlled application of the maps. Finding an answer to this question is the main goal of our paper.

\subsection{Daemonic ergotropy}\label{QSM_DaemErgotropy}

Having introduced the notation and the key figure of merit for our investigation, from here on we will assume the control qubit $Q$ to be in a pure initial state $\ket{\phi,\alpha} = \sqrt{\phi}\ket{0} + e^{i\alpha}\sqrt{1-\phi}\ket{1}$, with $\phi \in [0,1]$ and $\alpha\in[0,2\pi]$. We stress that, although a more general initial preparation can be considered, it is reasonable to adopt this working assumption, since by definition the control qubit $Q$ represents a controlled degree of freedom. Upon introducing the projector $\hat{\pi}_{\phi,\alpha} = |\phi, \alpha\rangle \langle\phi, \alpha| \equiv \hat{\rho}_Q$, the action of Eq.~(\ref{QSmap}) gives 
\begin{equation}
\begin{aligned}
    \hat{\rho}_{SQ}^{out} &= \mathbb{M}_{\hat{\pi}_{\phi,\alpha}}(\mathcal{A}, \mathcal{B})[\hat{\rho}^{in}_S] \color{black} \\
    &= { \phi\Bigl((\mathcal{A} \circ \mathcal{B})[\hat{\rho}^{in}_S] \otimes  |0\rangle \langle 0|\Bigr)} \\
    & + {(1 - \phi)\Bigl((\mathcal{B} \circ \mathcal{A})[\hat{\rho}^{in}_S] \otimes  |1\rangle \langle 1|\Bigr)} \\
    & + {e^{-i\alpha}\sqrt{\phi(1-\phi)} \Bigl( \chi[\hat{\rho}^{in}_S] \otimes |0\rangle \langle 1| \Bigr) + {\rm h.c.},}
    \end{aligned}
\end{equation}
where $h.c.$ stands for the hermitian conjugate, and where $\chi[\hat{\rho}^{in}_S] = \sum_{ij} \hat{A}_i \hat{B}_j \hat{\rho}^{in}_S \hat{A}_i^\dagger \hat{B}_j^\dagger$ is the so-called cross-map, which emerges from and encapsulates the quantum coherence in the state $|\phi,\alpha\rangle$ of the control qubit $Q$. On the one hand, we can ignore the control qubit $Q$ (hence, tracing it out) and perform a cyclical process on the work medium regardless of the way the maps were applied to it. In this case, we ignore the coherence presented in $Q$ and extract work from the output state
\begin{equation}
\label{PartialTrace}
\begin{aligned}\hat{\rho}^{out, class}_S &=\operatorname{Tr}_Q \Bigl( \mathbb{M}_{\hat{\pi}_{\phi,\alpha}}(\mathcal{A}, \mathcal{B})[\hat{\rho}^{in}_S] \color{black} \Bigr) \\
&= \phi (\mathcal{A} \circ \mathcal{B})[\hat{\rho}^{in}_S] + (1-\phi) (\mathcal{B} \circ \mathcal{A})[\hat{\rho}^{in}_S],
\end{aligned}
\end{equation}
that realizes the probabilistic mixture of consecutive applications of the maps $\mathcal{A}[\hat{\rho}^{in}_S]$ and $\mathcal{B}[\hat{\rho}^{in}_S]$ in accordance with Eq.~\eqref{ClassicalControl} and which we will refer to as \textit{classically controlled output state} in what follows. In this case, the maximum amount of extractable work is simply given by the ergotropy (what we will refer to as \textit{classical})
\begin{equation}\label{daemonicGainComm}
    W^{class}(\hat{\rho}^{out}_{SQ}) \equiv W(\hat{\rho}^{out,class}_S), 
\end{equation}
corresponding to the maximal work that can be extracted from $S$ undergoing the classically controlled maps without acquiring information on it.

On the other hand, more generally, we can acquire information on the work medium and find with its help a more optimal cyclical process to extract work from $S$. This can be done by measuring the control qubit, and, in this case, one needs to take into account the effect that a measurement of $Q$ has on the ergotropy. In particular, by performing a projective measurement $\hat{\pi}_a$ on $Q$ and communicating the result, one can obtain -- with probability $p_a = \operatorname{Tr}(\hat{\pi}_a \hat{\rho}_{SQ}^{out})$ -- the post-measurement state of the system
\begin{equation}\label{CondState}
\hat{\rho}_{S|a}^{out} = \frac{1}{p_a} \operatorname{Tr}_Q(\hat{\pi}_a \hat{\rho}_{SQ}^{out}),    
\end{equation}
where $\hat\pi_a$ is an orthogonal projection operator in the Hilbert space of the control qubit. This naturally brings us towards the scheme put forward in Ref.~\cite{Francica2017} and extended in \cite{Bernards2019}, which quantifies and characterizes the gain in ergotropy associated with the acquisition of information on $S$. Such gain originates from the quantity dubbed {\it daemonic ergotropy} in light of the role played by the control qubit (which is akin to a Maxwell demon)
\begin{equation}
\label{DaemonicErg}
 W^D_{\hat{\pi}_a}(\hat{\rho}_{SQ}^{out}) = \sum_a p_a  W(\hat{\rho}_{S|a}^{out}),
\end{equation}
that can be optimized by taking a maximum over the set of all projective measurements of the control qubit, $W^D(\hat{\rho}_{SQ}^{out}) = \max_{\{\hat{\pi}_a\}} W^D_{\hat{\pi}_a}(\hat{\rho}_{SQ}^{out})$.

In what follows, we aim to characterize the daemonic ergotropy associated with the output state $\hat{\rho}_{SQ}^{out}$ of the QSM and explore the role of coherence in the control qubit $Q$, which in our scheme reflects into a non-consecutive application of the two above maps $\mathcal{A}$, $\mathcal{B}$. Applying the daemonic ergotropy scheme to the QSM, we perform a measurement of the control qubit in a certain basis, e.g., the one spanned by
\begin{equation}
\begin{aligned}\label{GenBasis}
\ket{\phi',\alpha'}_+&=\sqrt{\phi'}\ket{0}+e^{i\alpha'}\sqrt{1-\phi'}\ket{1}, \\
\ket{\phi',\alpha'}_-&= -e^{-i\alpha'}\sqrt{1-\phi'}\ket{0}+\sqrt{\phi'}\ket{1}, 
\end{aligned}
\end{equation}
with $\phi' \in [0,1]$ and $\alpha'\in[0,2\pi]$. According to Eq.~\eqref{CondState}, this leads to the conditional states
\begin{equation}\label{Eq.general}
\begin{aligned}
   p_+ \hat{\rho}^{out}_{S|\phi',\alpha',+} &=  \phi \phi' (A {\circ} B)[\hat{\rho}^{in}_S] + (1-\phi)(1-\phi') (B {\circ} A)[\hat{\rho}^{in}_S] \\
   &+ \sqrt{\phi\phi'(1-\phi)(1-\phi')} \Bigl( e^{-i(\alpha - \alpha')} \chi[\hat{\rho}^{in}_S]+ {\rm h.c.} \Bigr), \\
p_- \hat{\rho}^{out}_{S|\phi',\alpha',-} &=  \phi(1-\phi') (A {\circ} B)[\hat{\rho}^{in}_S] + (1-\phi)\phi' (B {\circ} A)[\hat{\rho}^{in}_S] \\
   &- \sqrt{\phi\phi'(1-\phi)(1-\phi')} \Bigl( e^{-i(\alpha - \alpha')} \chi[\hat{\rho}^{in}_S]+ {\rm h.c.} \Bigr),
\end{aligned}
\end{equation}
where
\begin{equation}
\begin{aligned}
p_\pm &= \frac{1}{2} \pm \frac{1}{2} \Bigl((2\phi - 1)(2\phi' - 1) \\
&+ 2\sqrt{\phi\phi'(1-\phi)(1-\phi')} \operatorname{Tr}\Bigl[e^{-i(\alpha - \alpha')}\chi[\hat{\rho}^{in}_S] + {\rm h.c.}\Bigr]\Bigr)
\end{aligned}
\end{equation}
are the probabilities to find $S$ in the corresponding conditional state. The daemonic ergotropy~(\ref{DaemonicErg}) of these states leads to the \textit{daemonic gain}
\begin{equation}
\delta W_{\phi', \alpha'}(\hat{\rho}^{out}_{SQ}) = W^D_{\hat{\pi}_{\phi', \alpha'}}(\hat{\rho}^{out}_{SQ}) - W^{class}(\hat{\rho}^{out}_{SQ})
\end{equation}
in extractable work compared with the corresponding classical ergotropy [cf. Eq.~(\ref{daemonicGainComm})]. As proven in Ref.~\cite{Francica2017}, the daemonic gain is always non-negative for any state of the ancilla (i.e., the control qubit) and projective measurement of it. Hence, QSM allows to extract more work from $S$ than a simple classical control that puts the maps into a probabilistic mixture. Moreover, the conditional states in Eq.~(\ref{Eq.general}) can be  represented as
\begin{equation}\label{CondStatesGain}
   \hat{\rho}^{out}_{S|\phi',\alpha', \pm} =  \frac{1}{2} \hat{\rho}^{out, class}_S  \pm \hat{G}
   \end{equation}
with
\begin{equation}
    \begin{aligned}
   \nonumber \hat{G} &= \Bigl(\phi' - \frac{1}{2}\Bigr)\Bigl((\phi - \frac{1}{2})\{\mathcal{A}, \mathcal{B}\}[\hat{\rho}^{in}_S]  + \frac{1}{2}[\mathcal{A}, \mathcal{B}][\hat{\rho}^{in}_S]\Bigr)  \\
   \label{GainMatrix} &+ \sqrt{\phi\phi'(1-\phi)(1-\phi')} \Bigl( e^{-i(\alpha - \alpha')} \chi[\hat{\rho}^{in}_S] + {\rm h.c.} \Bigr).
   \end{aligned}
\end{equation}
Here, $\{\mathcal{A}, \mathcal{B}\}[\hat{\rho}^{in}_S] = (\mathcal{A} \circ \mathcal{B})[\hat{\rho}^{in}_S] + (\mathcal{B} \circ \mathcal{A})[\hat{\rho}^{in}_S]$ and $[\mathcal{A}, \mathcal{B}][\hat{\rho}^{in}_S] = (\mathcal{A} \circ \mathcal{B})[\hat{\rho}^{in}_S] - (\mathcal{B} \circ \mathcal{A})[\hat{\rho}^{in}_S]$ are the anti-commutator and the commutator of the maps, respectively. This suggests that the daemonic gain can emerge in principle from the interplay of two mechanisms encoded in $\hat{G}$
\begin{enumerate}[(1)]
    \item The coherence in the control qubit, hence, non-separability of the application order of the maps,
    \item The acquisition of ``which-order'' information with respect to the non-commutativity of the maps and the imbalance in their separable combination.
\end{enumerate}
The above argument thus remarks that, even in the case where a classically controlled application of non-commutative maps is performed, a daemonic gain can be achieved when the control qubit $Q$ is measured in a titled basis. Both sources of the daemonic gain are balanced by $\phi'$, hence, it is possible to tune their contribution by choosing the appropriate projective measurement of $Q$. Indeed, the choice $\phi'=\frac{1}{2}$ plays a special role since, in this case, the daemonic gain can emerge only from the coherent control of maps' application order (and is maximal if $\alpha' = \alpha$). 

It can be noticed that the above expressions significantly simplify in the case of commutative maps, i.e., such that $\mathcal{A} \circ \mathcal{B} = \mathcal{B} \circ \mathcal{A}$. Even in this scenario, however, the measurement performed on the control qubit $Q$ plays an important role in extraction of work, in accordance with Eq.~\eqref{CondState} and Eq.~\eqref{DaemonicErg}. Notice first of all that the cross-map can be decomposed in the following form,
\begin{equation}\label{CrossMap}
    \chi[\hat{\rho}^{in}_S] = \pm (\mathcal{A} \circ \mathcal{B})[\hat{\rho}^{in}_S] + \frac{1}{2} \chi^{nc}_\mp [\hat{\rho}^{in}_S],
\end{equation}
where 
\begin{equation}\label{CommContrib}
    \chi^{nc}_\mp [\hat{\rho}^{in}_S] = \sum_{i,j} \hat{A}_i \hat{B}_j \hat{\rho}^{in}_S [\hat{B}_j, \hat{A}_i]_\mp^\dagger,
\end{equation}
and $[.\, , .]_\mp$ refers to a commutator or an anticommutator, respectively. This means that if the Kraus operators belonging to different maps all commute or all anticommute, i.e., $\hat A_i \hat B_j = \pm \hat B_j\hat A_i $, then $\chi^{nc}_\mp [\hat{\rho}^{in}_S] = 0$, and the output of the cross-map $\chi[\hat{\rho}^{in}_S]$ becomes proportional to the contribution from classical control $\hat{\rho}_S^{out,class} = (\mathcal{A} \circ \mathcal{B})[\hat{\rho}^{in}_S]$. In other words, a measurement performed on the control qubit produces the same state $(\mathcal{A} \circ \mathcal{B})[\hat{\rho}^{in}_S]$ as if the control qubit was discarded. This allows us to draw our first important conclusion, namely that the daemonic ergotropy for maps $\mathcal{A}$ and $\mathcal{B}$ with mutually commutative or anti-commutative Kraus operators, is equal to the ergotropy for a consecutive application of the maps [cf. Eq.~(\ref{SepChan})]. Hence, the observation of an extra gain in the case of commutative maps witnesses the presence of quantum signatures in a thermodynamic system. While this might remind of similar considerations that could be made in regard to contextuality-related issues [cf. Ref.~\cite{Lostaglio2020}], any potential link with our observation above should be carefully scrutinized.

\subsection{Gain conditions}\label{sec:GainConditions}

Since the QSM allows for a non-negative gain $\delta W_{\phi', \alpha'}$ in extracted work from the output state through the daemonic ergotropy scheme compared with the classical control, it is important to find a condition for $\delta W_{\phi', \alpha'} = 0$. Indeed, it is possible to show that no extra gain in ergotropy can be achieved by measuring the control qubit $Q$ if and only if the conditional states in Eq.~(\ref{CondStatesGain}) are passive with respect to the Hamiltonian~\cite{Francica2017}
\begin{equation}
\hat{H'} = \sum_k \epsilon_k | r_k^{class} \rangle \langle r_k^{class} |,
\end{equation}
where $\hat{\rho}^{out,class}_S = \sum_k r_k^{class} | r_k^{class} \rangle \langle r_k^{class} |$ denotes the spectral decomposition of the classical control output in Eq.~(\ref{PartialTrace}). Hence, it is enough to check whether the state in Eq.~(\ref{CondStatesGain}) satisfy the conditions (a) and (b) of passivity discussed at the Section~\ref{Ergotropy}.

Condition (a) requires commutation of $H'$ and the conditional states in Eq.~(\ref{CondStatesGain}), while condition (b) requires that the more energetic of such states (with respect to $\hat{H}'$) are the least populated. Except for two extreme cases (which will be discussed later on), such conditions are equivalent to the following set of statements
\begin{enumerate}[(A)]
    \item The conditional states in Eq.~(\ref{CondStatesGain}) mutually commute,
    \item When using the representation based on the eigenbasis of the classical control output $\rho_S^{out,class}$, the diagonal elements of the conditional states in Eq.~(\ref{CondStatesGain}) have to be ordered in the same way as the diagonal elements of $\rho_S^{out,class}$. Thus, if the eigenvalues $r_k^{class}$ are such that $r_k^{class} \geq r_{k+1}^{class}$, then $(\hat{\rho}^{out}_{S|\phi',\alpha', \pm})_{kk} \geq (\hat{\rho}^{out}_{S|\phi',\alpha', \pm})_{k+1,k+1}$.
\end{enumerate}
Putting it all together and sorting $r_k^{class}$ in descending order, we obtain that $\delta W_{\phi', \alpha'} = 0$ if
\begin{widetext}
\begin{subequations}
\begin{eqnarray}
\label{GainCondition}
\sqrt{\frac{\phi'(1-\phi')}{\phi(1-\phi)}} \Bigl[\hat{\rho}_S^{out,class},\, e^{-i(\alpha-\alpha')} \chi[\hat\rho^{in}_S] + {\rm h.c.}\Bigr] &-& (2\phi'-1)\Bigl[(\mathcal{A} \circ \mathcal{B})[\hat{\rho}^{in}_S], (\mathcal{B} \circ \mathcal{A})[\hat{\rho}^{in}_S]\Bigr]=0, \\
2|G_{kk} - G_{k+1,k+1}| &\leq& r_{k}^{class} - r_{k+1}^{class},
\label{GainCondition2}
\end{eqnarray}
\end{subequations}
\end{widetext}
where $G_{kk}$ are the diagonal elements of $\hat{G}$ in the eigenbasis of Eq.~(\ref{PartialTrace}). These conditions put a constraint to the connection between the classically-controlled output and the contribution of quantum coherence presented in the control qubit. In turn, if we perform a measurement in the basis with $\phi' = {1}/{2}$, so that only the coherent control of the maps contributes to the daemonic ergotropy, Eqs.~(\ref{GainCondition})-(\ref{GainCondition2}) are significantly simplified to become
\begin{equation}
\begin{aligned}
\label{GainConditionCausal12}
[\hat{\rho}_S^{out,class},\, e^{-i(\alpha-\alpha')} \chi[\hat\rho^{in}_S] + {\rm h.c.}] &= 0, \\
 2\sqrt{\phi(1-\phi)}\Bigl|\operatorname{Re}[e^{-i(\alpha-\alpha')}\delta\chi_{kk}]\Bigr| &\leq \delta r^{class}_{k},
\end{aligned}
\end{equation}
where $\delta r^{class}_{k} = r_{k}^{class} - r_{k+1}^{class}$, and $\delta\chi_{kk} = \chi[\hat{\rho}^{in}_S]_{kk} - \chi[\hat{\rho}^{in}_S]_{k+1,k+1}$ are the differences of consequent diagonal elements of the cross-map in the eigenbasis of $\hat{\rho}_S^{out,class}$.
As mentioned above, there are two extreme cases that demand revising Eqs.~(\ref{GainCondition})-(\ref{GainCondition2}) and have thus to be studied separately. Firstly, it is a particular case of the fully degenerate classical control output, i.e., if it is a maximally mixed state $\hat{\rho}_S^{out,class} = {\mathds{1}}/{d}$, where $d$ is the dimension of the Hilbert space of $S$, and no ergotropic work can be extracted from it. In this case, any basis can be its eigenbasis: condition (a) is thereby automatically fulfilled since the eigenbasis of one of the conditional states can be taken as the eigenbasis of $\hat{\rho}_S^{out,class}$. On the other hand, in the light of condition~(\ref{GainCondition}), the fully degenerate $\rho_S^{out,class}$ means that $\delta r^{class}_{k} = 0$ for all $k$. Hence, in this case $\delta W_{\phi', \alpha'} = 0$ if and only if $\hat{G}$ is also fully degenerate,
\begin{equation}
  G_{kk} - G_{k+1,k+1} = 0.
\end{equation}
Hence, any bias between the populations of the energetic states brought by the quantum control of such maps leads to a non-zero value of work that can be extracted from their output (whereas the classical control does not allow to extract it at all). An example of such maps is given by completely depolarizing maps which are discussed in Section~\ref{Identmaps}. Another extreme case is obtained when the Hamiltonian of the system is fully degenerate, $\hat{H} = \epsilon \mathds{1}$. In this case, $\hat{H}' = \epsilon \mathds{1}$ is fully degenerate too, and the condition (a) is automatically fulfilled. On the other hand, the condition (b) is automatically fulfilled too since we can always take the eigenbasis of the corresponding conditional state with eigenvalues sorted in the descending order as the eigenbasis of $\hat{H}'$. Hence, $\delta W _{\phi', \alpha'} = 0$, which is intuitively clear: every state of $S$ has the same energy, and no cycle can decrease the internal energy of the system.

\section{Role of the quantum coherence in the work medium}

Interestingly, quantum coherence present in the state $\hat{\rho}_S$ of a system can be seen as a resource for work extraction, and its contribution to ergotropy can be meaningfully separated~\cite{Francica2020}. In fact, we can imagine the optimal unitary cycle $\hat{U}_c$, which outputs the passive state, as two transformations following each other. Firstly, we perform an incoherent transformation rearranging the energetic populations $\rho_{kk}$ in descending order (fulfilling thus the condition (b) of passivity). Then we continue with a cycle consuming the available coherence and obtaining the passive state. Ergotropy is thereby split into two contributions,
\begin{equation}\label{ErgSeparation}
W(\hat{\rho}_S) = W_i(\hat{\rho}_S) + W_c(\hat{\rho}_S),
\end{equation}
where $W_{i}(\hat{\rho}_S)$ quantifies the maximum amount of work that can be extracted from $\hat{\rho}_S$ without altering its quantum coherence, while $W_c(\hat{\rho}_S)$ quantifies the coherence-consuming counterpart. Similarly to Eq.~\eqref{ErgSeparation}, daemonic ergotropy can also be expanded in terms of incoherent and coherent contributions as
\begin{equation}
\label{DaemonicErg1}
 W^D_{\phi', \alpha'}(\hat{\rho}_{SQ}^{out}) = W^D_{\phi', \alpha'; i}(\hat{\rho}_{SQ}^{out}) + W^D_{\phi', \alpha'; c}(\hat{\rho}_{SQ}^{out}),
\end{equation}
where $W^D_{\phi', \alpha'; j} = \sum_a p_a  W_j(\hat{\rho}_{S|a}^{out})$ with $j = i,c$, respectively. Hence, the origin of the advantage provided by the quantum control of maps can be also traced via the corresponding counterpart of the daemonic gain,
\begin{equation}
\label{DaemonicGainOrigin}
 \delta W_{\phi', \alpha'; j}(\hat{\rho}_{SQ}^{out}) = W^D_{\phi', \alpha'; j}(\hat{\rho}_{SQ}^{out}) - W_{j}(\hat{\rho}_{SQ}^{out}),
\end{equation}
where $j = i,c$.

\subsection{Ergotropy of a qubit}\label{QubitErgotropy}

In order to fix the ideas and provide as clear-cut results as possible, in what follows, we consider a two-level quantum system $S$, with Hamiltonian $\hat H_S={\rm Diag}[\epsilon_1,\epsilon_2]$, initially in the generic state
\begin{equation}
    \hat{\rho}_S = \begin{pmatrix} \rho_{11} & \rho_{12}\\ \rho_{12}^* & \rho_{22} \end{pmatrix},\qquad(\rho_{11}+\rho_{22}=1).
\end{equation}
As before, we assume that the energy eigenvalues $\epsilon_k$ are sorted in increasing order as $\epsilon_1<\epsilon_2$. Without loss of generality, we can re-scale such energies so that $\epsilon_2=1$ and $\epsilon_1=0$. In this case, following Ref.~\cite{Francica2020}, the incoherent and coherent contributions to the ergotropy stated in Eq.~(\ref{ErgSeparation}) can be expressed explicitly as
\begin{equation}\label{QubitErgotropyContributions}
\begin{aligned}
W_i(\hat{\rho}_S)&= \max\{0,\delta\rho\},\\
W_c(\hat{\rho}_S)&= \frac{1}{2}\left(\eta-\sqrt{\eta^2 - 4 |\rho_{12}|^2}\right),
\end{aligned}
\end{equation}
where $\eta=\sqrt{2P(\hat\rho_S)-1}$ is a function of the purity $P(\hat{\rho}_S)=\operatorname{Tr}(\hat{\rho}_S^2)$ of the state of the system, and $\delta\rho=\rho_{22} - \rho_{11}$ denotes the population imbalance of the state under scrutiny.

Let us first focus on the general conditions under which an incoherent daemonic gain is achieved. If the control qubit is ignored, the maps are applied in a classically controlled order~(\ref{ClassicalControl}), and the following work can be extracted incoherently,
\begin{equation}
W^{class}_i(\hat{\rho}^{out}_{SQ}) = \text{max}\{0,\delta\rho^{out}_{class}\},
\end{equation}
where $\delta\rho^{out}_{class} = \phi\delta\rho^{AB} + (1-\phi)\delta\rho^{BA}$ and $\delta\rho^{XY}=(\mathcal{X} \circ \mathcal{Y})[\rho^{in}_S]_{22}-(\mathcal{X} \circ \mathcal{Y})[\rho^{in}_S]_{11}$ are the population imbalances of the outputs of the classically controlled and fixed-ordered maps, respectively. 

On the other hand, if we measure the control qubit in the basis embodied by the states in Eq.~(\ref{GenBasis}), the system is transformed into one of the conditional states in Eq.~(\ref{Eq.general}). Herewith, we can acquire information on $S$ after the action of the maps allowing to achieve the incoherent daemonic gain
\begin{equation}\label{IncoherWorkGeneral}
\delta W_{\phi',\alpha'; \, i}(\hat{\rho}_{SQ}^{out}) = \text{max}\Bigl\{0, |\delta G| - \frac{|\delta\rho^{out}_{class}|}{2}\Bigr\},
\end{equation}
with
\begin{equation}
\begin{aligned}
\delta G &= \frac{1}{2} (2\phi'-1) \Bigl( \phi\delta\rho^{AB} - (1-\phi)\delta\rho^{BA} \Bigr) \\
&+ 2\sqrt{\phi\phi'(1-\phi)(1-\phi')} \operatorname{Re}[e^{-i(\alpha - \alpha')} \zeta],
\end{aligned}
\end{equation}
where $\zeta = \chi[\hat{\rho}^{in}_{S}]_{22} - \chi[\hat{\rho}^{in}_{S}]_{11}$ is the imbalance of the diagonal elements of the cross-map in the energetic basis. Hence, if we extract work from $S$ via incoherent operations only, the QSM provides an advancement in ergotropy if and only if the maps satisfy the condition $|\delta\rho^{out}_{class}| < 2|\delta G|$. This implies that, in order to be thermodynamically advantageous, the QSM should increase the bias between the populations of the energy eigenstates of the Hamiltonian of the system $S$, in line with an intuitive expectation based on the physical interpretation of the incoherent work. The choice of an optimal projective measurement of the control qubit which makes the QSM maximally advantageous in incoherent work extraction, generally speaking, depends on the specific realization of the maps $\mathcal{A}$ and $\mathcal{B}$ and their action on $S$. This is the case when the control qubit is measured in the basis with $\alpha'=\alpha - \operatorname{arg}(\zeta)$ and
\begin{equation}
    \phi' = \frac{1}{2}\Bigl( 1 + \frac{\phi\delta\rho^{AB} - (1-\phi)\delta\rho^{BA}}{\sqrt{(\phi\delta\rho^{AB} - (1-\phi)\delta\rho^{BA})^2 + 4\phi(1-\phi)|\zeta|^2}} \Bigr),
\end{equation}
and, in turn, we achieve the daemonic gain
\begin{widetext}
\begin{equation}
\begin{aligned}
    \delta W_i(\hat{\rho}^{out}_{SQ}) &= \frac{1}{2} \text{max}\Biggl\{ 0, \sqrt{\Bigl( \phi\delta\rho^{AB} - (1-\phi)\delta\rho^{BA} \Bigr)^2 + 4\phi(1-\phi) |\zeta|^2}- |\delta\rho^{out}_{class}| \Biggr\}\\
    &= \frac{1}{2} \text{max}\Biggl\{ 0, \sqrt{\Bigl( |\delta\rho^{out}_{class}| - 2 \operatorname{sign}(\delta\rho^{AB}) \operatorname{sign}(\delta\rho^{BA}) \text{min}\Bigl\{\phi |\delta\rho^{AB}|, (1-\phi) |\delta\rho^{BA}|\Bigr\} \Bigr)^2 + 4\phi(1-\phi) |\zeta|^2}- |\delta\rho^{out}_{class}| \Biggr\}.
\end{aligned}
\end{equation}
\end{widetext}
Herewith, it is easy to identify the two sources (which are incoherent and coherent, respectively, with regard to the control qubit) of the daemonic gain mentioned in Section~\ref{sec:GainConditions}. On the one hand, non-commutativity of the maps can potentially enhance the extractable work if we are able to identify the application order of the maps, either $(\mathcal{A} \circ \mathcal{B})$ or $(\mathcal{B} \circ \mathcal{A})$. Indeed, this is the case when the output states $(\mathcal{A} \circ \mathcal{B})[\hat{\rho}_S^{in}]$ and $(\mathcal{B} \circ \mathcal{A})[\hat{\rho}_S^{in}]$ carry energetic population imbalances of opposite signs, i.e., the work can be extracted via incoherent operations only from one of them. In this case, acquiring the ``which order'' information, we can apply an incoherent cycle when the application order with a positive population imbalance in the output state is identified, otherwise ignore the output. In such a way, we can gain more work in comparison with a cycle that is applied always reshuffling thereby the average energetic populations with respect to the classically controlled output $\hat{\rho}_S^{out, class}$. Eventually, this contribution is compatible with an underlying causal structure and can be separated as ``causal'' incoherent daemonic gain
\begin{equation}
\begin{aligned}
\label{CausalIncoherGain}
    \delta W_{i, causal}(\hat{\rho}^{out}_{SQ}) &=\frac{1}{2}\Bigl(1-\operatorname{sign}(\delta\rho^{AB}) \operatorname{sign}(\delta\rho^{BA})\Bigr) \\
    &\qquad \cdot \text{min}\Bigl\{\phi |\delta\rho^{AB}|, (1-\phi) |\delta\rho^{BA}|\Bigr\}.
\end{aligned}
\end{equation}
The rest of the daemonic gain arises from the coherence in the control qubit which puts the fixed application orders $(\mathcal{A} \circ \mathcal{B})$ or $(\mathcal{B} \circ \mathcal{A})$ into a superposition. Hence, it can be seen as an extra resource for incoherent daemonic ergotropy provided by the QSM.

The causal incoherent daemonic gain arises from the non-commutativity of the maps, when the opposite application orders of them act on the work medium in different ways. If the maps are commutative, so that $\delta\rho^{out}_{class} = \delta\rho^{AB}$, the resulting daemonic gain arises exclusively from the coherence present in the control qubit,
\begin{widetext}
\begin{equation}
    \delta W_i(\hat{\rho}^{out}_{SQ}) = \frac{1}{2} \text{max}\Biggl\{ 0, \sqrt{(2\phi-1)^2(\delta\rho^{out}_{class})^2 + 4\phi(1-\phi) |\zeta|^2} - |\delta\rho^{out}_{class}| \Biggr\},
\end{equation}
\end{widetext}
and is maximal if the application orders are perfectly balanced by the control qubit, i.e., $\phi = 1/2$. In the following Subsections, we will consider several examples of the maps that admit the maximal daemonic gain for the control-qubit state $\ket{1/2,0}$ and projective measurements over the optimal basis with $\phi' = 1/2$ and $\alpha' = \alpha$ (since, in these examples, $\zeta \in \mathbb{R}$) which we will denote as $\{\ket{+}, \ket{-}\}$.

While the daemonic gain is always non-negative for incoherent operations, this is not true if the applied thermodynamic cycle consumes coherence present in the work medium,
\begin{widetext}
\begin{equation}
\begin{aligned}
    \delta W_{\phi', \alpha'; \, c}(\hat{\rho}^{out}_{SQ}) &= \frac{1}{2} \Biggr\{ \frac{1}{2} \sqrt{(\delta\rho^{out}_{class} + 2\delta G)^2 + 4|\rho^{out, class}_{12} + 2G_{12}|^2} +  \frac{1}{2} \sqrt{(\delta\rho^{out}_{class} - 2\delta G)^2 + 4|\rho^{out, class}_{12} - 2G_{12}|^2} \\
    &\qquad - \sqrt{(\delta\rho^{out}_{class})^2 + 4|\rho^{out, class}_{12}|^2} - \Bigl( \frac{1}{2}|\delta\rho^{out}_{class} + 2\delta G| + \frac{1}{2}|\delta\rho^{out}_{class} - 2\delta G| - |\delta\rho^{out}_{class}| \Bigr) \Biggr\},
\end{aligned}
\end{equation}
\end{widetext}
since the conditional states, in principle, can carry less coherence than the classically controlled output. Indeed, the expression between parentheses is the incoherent daemonic gain $\delta W_{\phi',\alpha'; \, i}(\hat{\rho}_{SQ}^{out})$, which means that the rest can be identified with the total daemonic gain in Eq.~(\ref{DaemonicGainOrigin}). Hence, $\delta W_{\phi', \alpha'; \, c}(\hat{\rho}^{out}_{SQ})$ has a sophisticated behavior which crucially depends on the action of the maps $\mathcal{A}$ and $\mathcal{B}$ on the coherence of the system's state. We leave thus the substantive analysis of the coherent counterpart of the daemonic gain for the following sections, where the specific realizations of the maps $\mathcal{A}$ and $\mathcal{B}$ are considered. Nevertheless, it can be shown that for the maps whose classically controlled output  $\hat{\rho}^{out,class}$ does not carry any coherence the coherent daemonic gain $\delta W_{\phi',\alpha'; \, c}(\hat{\rho}_{SQ}^{out}) \neq 0$ if and only if $G_{12} \neq 0$. On the other hand, it can happen the daemonic gain can be achieved only by consuming coherence of the output state, so that $\delta W_{\phi',\alpha'}(\hat{\rho}_{SQ}^{out}) = \delta W_{\phi',\alpha'; \, c}(\hat{\rho}_{SQ}^{out})$: this is the case when $|\delta\rho^{out}_{class}| \geq 2|\delta G|$ and $\delta W_{\phi',\alpha'; i}(\hat{\rho}_{SQ}^{out}) = 0$. Then, if the classically controlled application of the maps do not output a maximally mixed state, $\hat{\rho}^{out,class} \neq \frac{\mathds{1}}{d}$ (i.e., $\delta\rho^{out}_{class}$ and $\rho_{12}^{out, class}$ are not simultaneously equal to zero), the coherent daemonic gain (as well as the entire daemonic gain) is zero if the maps fulfill the condition
\begin{equation}
\delta\rho^{out}_{class}  G_{12} = \delta G  \rho_{12}^{out, class} 
\Rightarrow \delta W_{\phi',\alpha'}(\hat{\rho}_{SQ}^{out}) = 0.
\end{equation}
In the extreme case of a maximally mixed classically controlled output state $\hat{\rho}^{out,class} = {\mathds{1}}/{d}$ the corresponding condition reads $\delta G = G_{12} = 0$, as it can be seen from the example of completely depolarizing maps discussed in the following Subsection.

\subsection{Identical maps: Complete depolarization}\label{Identmaps}

While the considerations above refer to generic maps, it is informative to address explicitly exemplary cases. First we consider a simple case of $\mathcal{A}$ and $\mathcal{B}$ being not only commutative but identical maps, i.e., $\mathcal{A}[\hat{\rho}^{in}_S] = \mathcal{B}[\hat{\rho}^{in}_S]$. Interestingly, even this scenario can lead to a thermodynamically non trivial result in terms of daemonic ergotropy gain. In light of the formalism put forward in Sec.~\ref{Ergotropy}, in fact, whenever the map $\mathcal{A}[\hat{\rho}^{in}_S]$ has non-commutative Kraus operators, a classically controlled application of two identical copies of it leads to a finite daemonic gain according to Eq.~\eqref{DaemonicErg}.

In order to better illustrate this point, let us consider the explicit example of a completely depolarizing map $\mathcal{D}[\hat{\rho}^{in}_S]$. Since the output of the latter is a maximally mixed state, i.e.,
\begin{equation}\label{DepolChanQubit}
\mathcal{D}[\hat{\rho}^{in}_S]=\frac{1}{4}\sum_{i=1}^{4}\hat{U}_i \hat{\rho}^{in}_S \hat{U}_i^\dagger = \frac{\mathds{1}}{2}, 
\end{equation}
with $\hat{U}_i$ being a set of orthogonal unitary operators, it is a passive state, and no ergotropic work can be extracted from a system sent through such a map. This further implies that, seen as a communication channel, it cannot be used for effective information transmission. However, if the occurrence of each map's copy is controlled by the control qubit, then some communication and thermodynamic tasks such as classical information communication~\cite{Ebler2018} and cooling cycles~\cite{Felce2020} can be effectively performed. In the QSM, we have $\mathcal{A} = \mathcal{B} = \mathcal{D}$ with $(\mathcal{A} \circ \mathcal{B})[\hat{\rho}^{in}_S] = \mathcal{D}[\hat{\rho}^{in}_S]$. A straightforward calculation leads to $\chi[\hat{\rho}^{in}_S] = {\hat{\rho}^{in}_S}/{4}$ for the the cross-map in Eq.~\eqref{CrossMap} and to
\begin{equation}
p_\pm \cdot \hat{\rho}^{out}_{S|\pm} = \frac{\mathds{1}}{4} \pm \frac{\hat{\rho}^{in}_S}{8},
\end{equation}
for the un-normalized conditional states after the measurement of the control qubit onto the basis $\{\ket{+}, \ket{-}\}$. Since the output state in Eq.~(\ref{DepolChanQubit}) is maximally mixed, no ergotropic work can be extracted from it. This means that, for a classically controlled occurrence of the copies of $\mathcal{D}[\hat{\rho}^{in}_S]$, the corresponding ergotropy in Eq.~(\ref{daemonicGainComm}) is zero, i.e., $W^{class}(\hat{\rho}^{out}_{SQ}) = 0$, and the gain in ergotropy is simply the daemonic ergotropy, $\delta W = W^D$.  In particular, the incoherent contribution Eq.~(\ref{IncoherWorkGeneral}) to the daemonic ergotropy coming from the QSM is non-zero and equal to
\begin{equation}
W^D_i(\hat{\rho}^{out}_{SQ}) = {\frac{{|\delta\rho|}}{8}},
\end{equation}
i.e., proportional to the population imbalance $\delta\rho$ of the system's initial state $\hat{\rho}^{in}_S$. On similar footing, the coherent contribution to the daemonic ergotropy is also finite and reads
\begin{equation}
W^D_c(\hat{\rho}^{out}_{SQ}) = \frac{1}{8}\Bigl( \sqrt{(\delta\rho)^2 + 4|\rho_{12}|^2} - |\delta\rho|\Bigr).
\end{equation}
We furthermore see that this term is always non-negative and, as expected, non-zero whenever $\rho_{12}\neq 0$. Combining these results, we find that the total daemonic gain for the state sent through two copies of $\mathcal{D}[\hat{\rho}^{in}_S]$ controlled by $Q$ is given by
\begin{equation}
W^D(\hat{\rho}^{out}_{SQ}) = \frac{1}{8}\sqrt{(\delta\rho)^2 + 4|\rho_{12}|^2}.
\end{equation}
This physically means that a quantum-controlled occurrence of two identical completely depolarizing maps allows for an effective work extraction if the state of the system has a non-zero coherence or if it has a population bias, i.e., $\delta\rho\neq 0$. Notice that for the maximal coherence of $\hat{\rho}^{in}_S$, i.e., $\rho_{12}=\sqrt{\rho_{11}\rho_{22}}$, one obtains the values
\begin{equation}
\begin{aligned}
W^D_c(\hat{\rho}^{out}_{SQ}) &= \frac{1}{8}\Bigl( 1 - {|\delta\rho|}\Bigr) = \frac{1}{4}\min\{\rho_{11},\rho_{22}\}, \\
W^D(\hat{\rho}^{out}_{SQ}) &= \frac{1}{8},
\end{aligned}
\end{equation}
so that the total gain in ergotropy is the same for any pure state of the system.

The results for the depolarizing channel in Eq.~(\ref{DepolChanQubit}) can be easily generalized to a $d$-dimensional system with Hamiltonian $\hat H_S={\rm Diag}[\epsilon_1, ..., \epsilon_d]$, where the action is described by the map 
\begin{equation}
\mathcal{D}[\hat{\rho}^{in}_S]=\frac{1}{d^2}\sum_{i=1}^{d^2}\hat{U}_i \hat{\rho}^{in}_S \hat{U}_i^\dagger = \frac{\mathds{1}}{d}. 
\end{equation}
Clearly, if the maps' occurrence is controlled classically, we obtain $(\mathcal{A} \circ \mathcal{B})[\hat{\rho}^{in}_S] = \frac{\mathds{1}}{d}$ and $W^{class}(\hat{\rho}^{out}_{SQ}) = 0$. Conversely, for the pure state $\ket{\phi,\alpha}$ of the control qubit $Q$, we have that
\begin{equation}
\hat{\rho}^{out}_{S|\pm} = \frac{\mathds{1}}{2d} \pm \frac{\hat{\rho}_S}{2d^2}.
\end{equation}
The incoherent contribution is
\begin{equation}
W^D_i(\hat{\rho}^{out}_{SQ}) = \frac{1}{2d^2} \sum_k \epsilon_k (\rho^>_{kk}-\rho^<_{kk}), 
\end{equation}
where $\rho^<_{kk}$ ($\rho^>_{kk}$) are the populations of $\rho^{in}_S$ in decreasing (increasing) order and the energies $\epsilon_k$ {are understood to be arranged in increasing order}. If there is coherence among the energy eigenstates, then the gain will be increased by the amount $W^D_c>0$ given by
\begin{equation}
W^D(\hat{\rho}^{out}_{SQ}) = \frac{1}{2d^2}\sum_k \epsilon_k (r^>_k-r^<_k),
\end{equation}
where $r^<_k$ ($r^>_k$) are the eigenvalues of $\rho^{in}_S$ in decreasing (increasing) order.
For a pure state, this reduces to
\begin{equation}
W^D(\hat{\rho}^{out}_{SQ})=\frac{\epsilon_d-\epsilon_1}{2d^2}.
\end{equation}

\subsection{Identical maps: Thermalization}\label{identMapsTherm}

In the thermodynamic context, the role of ``depolarizing'' maps is played by thermalizing maps $\mathcal{T}_T[\hat{\rho}_S^{in}]$ that describe the process of reaching thermal equilibrium with a reservoir at the temperature $T$. Indeed, such a map outputs a Gibbs (thermal) state 
\begin{equation}
\hat{\rho}^T = \frac{1}{d} \hat{A} \sum_{i=1}^{d^2}\hat{U}_i \hat{\rho}^{in}_S \hat{U}_i^\dagger \hat{A}^\dagger = \frac{1}{Z}e^{-\beta\hat{H}},
\end{equation}
where $\beta = (kT)^{-1}$ is the inverse temperature, $Z = \operatorname{Tr}(e^{-\beta\hat{H}})$, and $\hat{A}$ is the square root of $\hat{\rho}^T$. Importantly, thermal states are unique \textit{completely passive} states: no ergotropic work can be extracted not only from $\hat{\rho}^T$ itself but even from a set $(\hat{\rho}^T)^{\otimes N}$ of $N$ its copies. A possible gain via the QSM is given hence again by the daemonic ergotropy itself, $\delta W = W^D$. The depolarizing map $\mathcal{D}$ can be seen then as a thermalizing map in the limit of an infinite temperature of the reservoir coupled to the system.

Now, we assume that the system $S$ is thermalized by two reservoirs at the same temperature $T$, i.e., $\mathcal{A} = \mathcal{B} = \mathcal{T}_T$. We return to the two-level $S$ and the corresponding Hamiltonian $\hat{H}_S = {\rm Diag}[0, 1]$ introduced in Sec.~\ref{QubitErgotropy}. The output of $\mathcal{T}_T[\hat{\rho}_S^{in}]$ and, in turn, the corresponding classically controlled output read thereby in the energetic basis
\begin{equation}\label{DepolChanQubit}
\hat{\rho}^{out, class}_S = \mathcal{T}_T[\hat{\rho}^{in}_S] = \frac{1}{1+e^{-\beta}} \begin{pmatrix} 1 & 0 \\ 0 & e^{-\beta} \end{pmatrix}, \end{equation}
On the other hand, if the placement of the reservoirs is controlled via the QSM, the following cross-map arises,
\begin{equation}\label{DeltaDepol}
    \chi[\hat{\rho}^{in}_S] = \hat{\rho}^T\hat{\rho}^{in}_S\hat{\rho}^T = \frac{1}{(1+e^{-\beta})^2} \begin{pmatrix} \rho_{11} & e^{-\beta} \rho_{12} \\ e^{-\beta} \rho_{12}^* & e^{-2\beta}\rho_{22} \end{pmatrix}.
\end{equation}
In turn, it leads to the un-normalized conditional states
\begin{equation}
p_\pm \cdot \hat{\rho}^{out}_{S|\pm} = \frac{1}{2(1+e^{-\beta})} \Biggl[ \begin{pmatrix} 1 & 0 \\ 0 & e^{-\beta} \end{pmatrix} \pm \frac{1}{1+e^{-\beta}} \begin{pmatrix} \rho_{11} & e^{-\beta} \rho_{12} \\ e^{-\beta} \rho_{12}^* & e^{-2\beta}\rho_{22} \end{pmatrix} \Biggr], \end{equation}
arising after the measurement of the control qubit $Q$ onto the basis $\{\ket{+}, \ket{-}\}$. For certain population imbalances $\delta\rho$ of the initial state $\hat{\rho}^{in}$, the corresponding incoherent daemonic gain is non-zero,
\begin{equation}\label{ThermalIncoherErgotropy}
    W_i^D(\hat{\rho}^{out}_{SQ}) = \begin{cases}
    (1 + e^{-2\beta}) \delta\rho - 3(1 - e^{-2\beta}), &\delta\rho > 3\frac{1 - e^{-2\beta}}{1 + e^{-2\beta}}, \\
    -(1 + e^{-2\beta}) \delta\rho - (1 - e^{-2\beta}), &\delta\rho < -\frac{1 - e^{-2\beta}}{1 + e^{-2\beta}}, \\
    0, &\text{otherwise}.
    \end{cases}  
\end{equation}
In particular, the work medium $S$ can initially stay in equilibrium with another reservoir at the temperature $T^{in}$ and, hence, be in a thermal state $\hat{\rho}^{in} = \hat{\rho}^{T^{in}}$ before to undergo the action of the thermalizing maps controlled by the QSM. In this case, the entire daemonic ergotropy is given by its incoherent counterpart as the conditional states do not carry any coherence. In fact, we have
\begin{widetext}
\begin{equation}
p_\pm \hat{\rho}^{out}_{S|\pm} = \frac{1}{2(1+e^{-\beta})}  \begin{pmatrix} 1 \pm \frac{1}{(1+e^{-\beta})(1+e^{-\beta^{in}})} & 0 \\ 0 & e^{-\beta} \Bigl[1 \pm \frac{e^{-(\beta + \beta^{in})}}{(1+e^{-\beta})(1+e^{-\beta^{in}})}\Bigr] \end{pmatrix}. \end{equation}
\end{widetext}
Plugging in the initial population imbalance $\delta\rho = - \frac{1-e^{-\beta}}{1+e^{-\beta}}$ into Eq.~(\ref{ThermalIncoherErgotropy}), we obtain the daemonic gain
\begin{equation}
    W^D(\hat{\rho}^{out}_{SQ}) = \operatorname{max}\Bigl\{ 0, \frac{e^{-2\beta} - e^{-\beta^{in}}}{2(1+e^{-\beta})^2(1+e^{-\beta^{in}})}\Bigr\}.
\end{equation}
Hence, a non-zero ergotropic work can be extracted from a system which is thermalized via the QSM if
\begin{equation}
    T > 2T^{in},
\end{equation}
i.e., if the system is heated by the quantum-controlled reservoirs to a temperature more than two times its initial one.

\subsection{Non-identical maps: Amplitude damping and dephasing}\label{QubitGain}

We continue our analysis by generalizing the above scenario to non-identical maps, either describing a generalized amplitude damping or a phase-flip, whose composition can be regarded under certain conditions as a thermalizing map. The former is described by the following Kraus operators,
\begin{equation}
\begin{aligned}
& \hat{A}_0 = \sqrt{p} \begin{pmatrix} 1 & 0\\ 0 & \sqrt{1-\gamma} \end{pmatrix}, \; \hat{A}_1 = \sqrt{p} \begin{pmatrix} 0 & \sqrt{\gamma} \\ 0 & 0 \end{pmatrix}, \;\\
& \hat{A}_2 = \sqrt{1 - p} \begin{pmatrix} \sqrt{1-\gamma} & 0\\ 0 & 1 \end{pmatrix}, \; \hat{A}_3 = \sqrt{1 - p} \begin{pmatrix} 0 & 0 \\ \sqrt{\gamma} & 0 \end{pmatrix},
\end{aligned}
\end{equation}
while the latter by
\begin{eqnarray}
\hat{B}_0 = \sqrt{q} \openone,
\; \hat{B}_1 = \sqrt{1-q}\hat{\sigma}_z,
\end{eqnarray}
with $\openone$ being the identity matrix and $\hat{\sigma}_{k}~(k=x,y,z)$ the $k$-Pauli matrix. The parameters $\gamma,q\in[0,1]$ quantify the strength of the channel, while $p\in[0,1]$ determines the chance that incoherent damping (described by $\hat A_{0,1}$) or pumping (resulting from the application of $\hat A_{2,3}$) occur. Importantly, maps $\mathcal{A}$ and $\mathcal{B}$ commute, a fact which, in light of Eq.~\eqref{PartialTrace}, implies that an initial state $\hat\rho^{in}_S$ is mapped into
\begin{equation}
\label{mappa}
\begin{aligned}
(\mathcal{A} \circ \mathcal{B})[\hat{\rho}^{in}_S] &= (1-\gamma) \begin{pmatrix} \rho_{11} & 0 \\ 0 & \rho_{22} \end{pmatrix} + \gamma \begin{pmatrix} p & 0 \\ 0 & 1 - p \end{pmatrix}\\
&- (1 - 2q) \sqrt{1-\gamma} \begin{pmatrix} 0 & \rho_{12} \\ \rho_{12}^* & 0\end{pmatrix}.
\end{aligned}
\end{equation}
For $\gamma = 1$, $q=1/2$ or $\rho_{12}=0$, the state in Eq.~\eqref{mappa} has no quantum coherence. In turn, this means that the contribution to the classical ergotropy is fully incoherent $W^{class} =W^{class}_i$. In particular, $(\mathcal{A} \circ \mathcal{B})$ can be regarded as a thermalizing map if $\gamma = 1$ and $\frac{1}{2} \leq p \leq 1$~\cite{Guha2020}.

\begin{table}[b!]
\caption{The values of the population imbalance $x_\pm$, leading to the essentially coherent gain, in the ranges defined by the values taken by $p$.}
\begin{tabular}{c|c|c}
&$p<1/2$     & $p>1/2$  \\
\hline\hline
$x_+$ & $1-2p$     & $\dfrac{\gamma(1+q)|1-2p|}{2-\gamma(1+q)}$\\
\hline
$x_-$& $\dfrac{\gamma(1+q)(1-2p)}{2-\gamma(1+q)}$& $|1-2p|$
\end{tabular}
\label{tav}
\end{table}

Taking a generic state of the control qubit and assuming arbitrary projective measurements, we achieve Eq.~\eqref{Eq.general} with the following contribution from the quantum coherence in the state of the control qubit
\begin{equation}\label{DeltaComm}
   \chi[\hat{\rho}^{in}_S] = (\mathcal{A} \circ \mathcal{B})[\hat{\rho}^{in}_S] - \gamma(1-q)\begin{pmatrix} p(1 + \delta\rho) & 0 \\ 0 & (1-p)(1-\delta\rho) \end{pmatrix},
\end{equation}
which bears dependence on the population imbalance $\delta\rho$ of the initial state. Notice that if $\delta\rho = 1-2p$, we get the trivial case $\chi^{nc}_\mp[\hat{\rho}_S] = 0$ due to Eq.~(\ref{CommContrib}). This means that the populations and the eigenvalues of each conditional state will be shifted by the same amount, and the resulting gain $\delta W$ in ergotropy is zero.

The gain $\delta W$ is maximized by the choice of $\phi=1/2,\alpha=0$, which maximizes the initial coherence in the state of the control qubit. The optimal projective measurement is the one performed over the basis $\{\ket{+}, \ket{-}\}$. Due to Eq.~(\ref{DeltaComm}), the off-diagonal elements in the output state originate solely from the action of the maps $(\mathcal{A} \circ \mathcal{B})[\hat{\rho}_S]$ themselves. Hence, only the conditional state 
\begin{equation}\label{PlusCondState}
\begin{aligned}
p_+ \cdot \hat{\rho}^{out}_{S|+} &= (\mathcal{A} \circ \mathcal{B})[\hat{\rho}_S] \\
&- \frac{\gamma}{2}(1-q)\begin{pmatrix} p(1 + \delta\rho) & 0 \\ 0 & (1-p)(1-\delta\rho)\end{pmatrix}
\end{aligned}
\end{equation}
obtained by projecting upon $\ket{+}$ bears coherences, whereas 
\begin{equation}
{p_- \cdot \hat{\rho}^{out}_{S|-} = \frac{\gamma}{2}(1-q)\begin{pmatrix} p(1 + \delta\rho) & 0 \\ 0 & (1-p)(1-\delta\rho)\end{pmatrix}}
\end{equation}
is diagonal since the action of $(\mathcal{A} \circ \mathcal{B})[\hat{\rho}_S]$ is canceled out. In this way, solely Eq.~(\ref{PlusCondState}) would contribute to the coherent part $W^D_c(\hat{\rho}_S)$ of daemonic ergotropy. On the other hand, due to Eq.~(\ref{mappa}), for a maximal strength of the amplitude damping channel $\mathcal{A}$ ($\gamma = 1$), a balanced phase flip channel $\mathcal{B}$ ($q=1/2$) or incoherent state $\hat{\rho}_S$ of the system ($\rho_{12}=0$), the corresponding daemonic ergotropy has a purely incoherent nature, and, thus, $W^D = W^D_i$.

Focusing on the incoherent counterpart [Eq.~(\ref{IncoherWorkGeneral})] of the ergotropy, we obtain
\begin{eqnarray}
   \delta\rho^{out}_{class} &=& \gamma (1-2p) + (1-\gamma)\delta\rho, \\
   {\zeta} &=& \delta\rho^{out}_{class} + \gamma (1-q) \left[\delta\rho - (1 - 2p) \right].
\end{eqnarray}
This leads to the following condition that should be met in order to achieve incoherent gain
\begin{equation}
    \Bigl|  \delta\rho^{out}_{class} + \gamma (1-q) [{\delta\rho} - (1 - 2p)] \Bigr| > |\delta\rho^{out}_{class}|,
\end{equation}
which entails $\delta\rho \notin [-x_-, x_+]$, where we have introduced the population imbalances $x_\pm$, whose expressions against the values taken by $p$ are given in Table~\ref{tav}. In particular, a more effective ergotropic work extraction can be performed for the initially passive states with $\delta\rho < -x_-$ being sent through the quantum switch. Notice that for $\gamma=0$ ($q=1$), channel $\mathcal{A}$ ($\mathcal{B}$) reduces to an identity map, whose Kraus operators commute with any other operator, so that $\chi^{nc}_\mp[\hat{\rho}_S] = 0$ due to Eq.~(\ref{CommContrib}). Hence, coherence of the control qubit plays no role in this case, and no gain can be achieved.

\begin{figure}[!b]
\includegraphics[width=0.99\columnwidth]{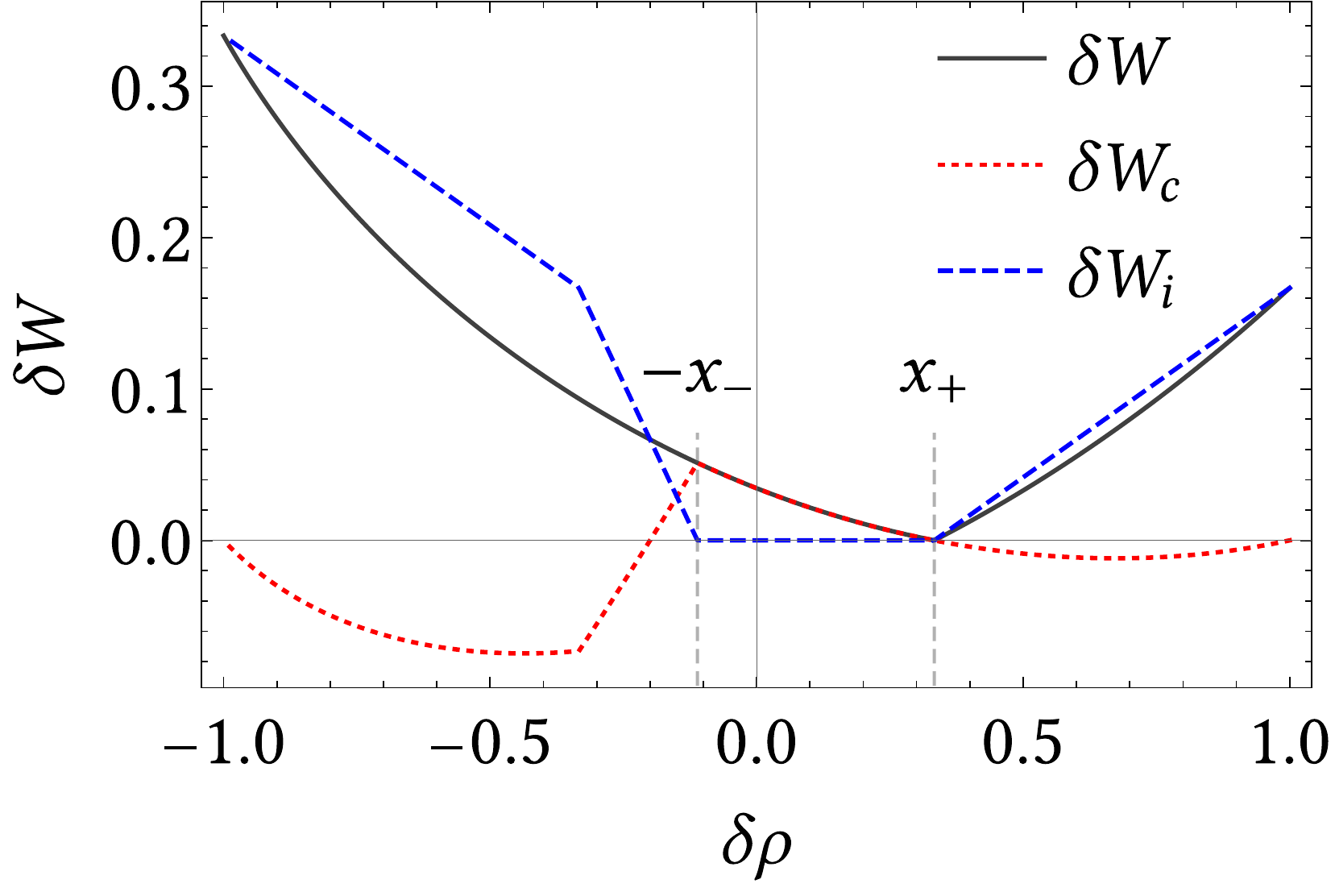}
\caption{We plot the gain $\delta W$ and its coherent $\delta W_c$ and incoherent $\delta W_i$ contributions against the population imbalance $\delta\rho$. We have taken $\gamma=1/2$, $p=1/3$ , $q=0$ and $\rho_{12}=\sqrt{\rho_{11} \rho_{22}}$, i.e., the state $\hat{\rho}_S$ has the maximal coherence and, hence, is pure. Notice that the gain is fully coherent in the interval $\delta\rho\in[-x_-,x_+]$. The energy scale and, therefore, the scale of the ergotropic gain $\delta W$ is given in unit of $\epsilon_2$ with the zero chosen in a manner that $\epsilon_1 = 0$ is satisfied.}\label{fig:gain}
\end{figure}

We address the case of $q=0$, which corresponds to the strongest possible action of the phase-flip channel. For $\gamma < 1$ and $p\neq\frac{1}{2}$, when the strength of the amplitude damping channel $\mathcal{A}$ is not maximal, and the phase flip channel $\mathcal{B}$ is imbalanced, the quantum coherence in $\hat{\rho}_S$ can contribute to $\delta W$. In contrast to the incoherent counterpart, the coherent contribution $\delta W_c = W^D_c - W^{class}_c$ can be positive as well as negative. In particular, the gain in ergotropic work has an exclusively coherent origin if it is extracted from the states with the population imbalance lying in the interval $\delta\rho \in [-x_-, x_+]$. This means that, for $\rho_{12} = 0$, a non-zero gain $\delta W$ can be achieved only for the states with $\delta\rho \notin [-x_-, x_+]$. 

In the presence of coherence in $\hat{\rho}_S$ this condition reduces to $\delta\rho \neq 1-2p$, i.e., more work can be potentially extracted from the state $\hat{\rho}_S$ sent through the quantum switch if its population imbalance differs from one between the incoherent damping and pumping of channel $\mathcal{A}$. Otherwise, the amplitude damping channel does not change the relative population of the eigenstates in $\hat{\rho}_S$, so that $\chi^{nc}_\mp[\hat{\rho}_S] = 0$ due to Eq.~(\ref{CommContrib}), and no ergotropic gain $\delta W$ is acquired. For example, setting $p=1/2$ makes the incoherent damping and pumping of channel $\mathcal{A}$ equally important. In this case, quantum coherence in $\hat{\rho}_S$ contributes poorly to the ergotropic gain $\delta W$, which, thus, is produced mainly by the incoherent counterpart $\delta W_i = W^D_i - W^{class}_i$ and can be gained for the imbalanced states $\hat{\rho}_S$, i.e., $\delta\rho \neq 0$. On the other hand, there always exist at least two values of the initial population imbalance $\delta\rho$ which lead to the daemonic gain of exclusively incoherent origin, i.e., $\delta W = \delta W_i$ and $\delta W_c = 0$. At first, this is the case when $\delta\rho = x_+$, and the entire daemonic gain is zero. Another point is given by $\delta\rho = \frac{\gamma(2p-1)(3+q)}{4-\gamma(3+q)}$ which makes the classically controlled output and the conditional states carrying the same amount of coherence.

In Fig.~\ref{fig:gain} we represent the gain $\delta W$ with its incoherent and coherent contributions $\delta W_i = W^D_i-W^{class}_i$ and $\delta W_c$. We note that the coherent gain can be larger or smaller than the classical coherent contribution, and in the interval $\delta\rho \in [-x_-,x_+]$ the incoherent part is null, thus delivering a fully coherent gain. For $q\neq 0$ and $q\neq 1/2$ we have the same behavior until the parameters are such that $x_+>1$ (or $x_->1$), for which the gain goes to zero also at $\rho_{22}=1$ (or $\rho_{11}=1$) since both incoherent and coherent gains are zero.

\section{Conclusions and Outlook}\label{Conclusions}

The potential benefits of coherent activation of maps in thermodynamic tasks have attracted recently much of interest in the scientific community~\cite{Nie2020, Cao2021, Felce2021, Felce2020, Goldberg2021, Nie2022, Guha2020, Chen2021, Wood2021}. In this paper, we have investigated the advantages in work extraction from a finite quantum system entailed by the quantum switch model that puts the application order of the maps under coherent control. Specifically, we have applied to the quantum switch model the concept of daemonic ergotropy which takes into account the assistance from an ancillary system in work extraction via unitary cycles~\cite{Francica2017}. In this regard, the action of two maps in a quantum-controlled order and the following performance of a suitable projective measurement on the control qubit allow for a non-zero gain in ergotropy compared to a classically controlled occurrence of the same maps. Importantly, we formulated a requirement that the maps should satisfy to produce such gain putting in turn certain constraints on the contribution of coherence in the control qubit.

We have illustrated our findings within paradigmatic examples of maps whose application order is controlled via the quantum switch model. In the simplest scenario, two copies of a certain map act on the work medium. Indeed, double action of the map is crucial here: the system's state undergoes the action of two not necessarily commutative Kraus operators belonging to different copies of the map. Hence, the application order of the Kraus operators plays indeed a role, and its quantum control allows in principle for a non-zero gain in ergotropy. The completely depolarizing map and its thermodynamic counterpart, thermalizing map, are of particular interest since they output by construction a completely passive state which by no means allows to extract ergotropic work from the system. Nevertheless, two copies of such maps placed in a superposition of their application orders via the quantum switch model activate the system's state and allow thereby to extract non-zero work from it if an appropriate measurement of the control qubit is performed. This suggests a charging protocol for a quantum battery that can be charged, for example, by coupling it to two identical reservoirs in an order controlled via the quantum switch model. We have also extended our discussion by considering the case of non-identical quantum maps. In particular, we considered a combination of an amplitude damping and a phase-flip channel which can be regarded as a thermalizing map under certain conditions. A detailed study of the incoherent and coherent contributions to the daemonic ergotropy has shown that the incoherent daemonic gain is always non-negative, whereas the coherent one can be positive as well as negative depending on the initial state of the system. Moreover, we have provided conditions concerning the initial system state in order to lead to a purely coherent gain in ergotropy: for such a scenario, the quantum switch model still allows for a non-zero gain with respect to the classically-controlled application of the maps.

Importantly, activation of the system's state by the maps applied in a quantum-controlled order even when the input and classically-controlled output states are passive could be a hint for a thermodynamic analogue of the recently developed resource theory of communication with quantum-controlled order of channels~\cite{Kristjansson2020}. We expect such a resource theory to be richer than, on the one hand, its communication counterpart due to a larger set of the free states, and the existing resource theories of thermodynamics because of a larger class of allowed thermal operations. We also believe that the obtained results can be implemented experimentally, for example, via photonic setups recently used to demonstrate the advantages of quantum-controlled ordering of maps in thermodynamic protocols~\cite{Nie2020, Cao2021}. Hence, we hope that this work will stimulate the research on the applications of quantum-controlled maps in thermodynamics.

\section*{Acknowledgements}
We are grateful to \v{C}aslav Brukner, Giulio Chiribella, Philippe Allard Gu\'erin, Tamal Guha, Hl\'er Kristj\'ansson, Ilya Kull, Gonzalo Manzano, Giulia Rubino, Roy Saptarshi, Uttam Singh, and Zolt\'an Zimbor\'as for the fruitful discussions and three anonymous referees for the useful and valuable comments to the manuscript. We acknowledge support from the Austrian Science Fund (Grant No. FWF-P30821), the European Research Council Starting Grant ODYSSEY (Grant No. 758403), the H2020-FETOPEN-2018-2020 project TEQ (Grant No. 766900), the DfE-SFI Investigator Programme (Grant No. 15/IA/2864), COST Action CA15220, the Royal Society Wolfson Research Fellowship (RSWF/R3/183013) and International Exchanges Programme (IEC/R2/192220), the Leverhulme Trust Research Project Grant (Grant No. RGP- 2018-266), the UK EPSRC (EP/T028424/1), and the Department for the Economy Northern Ireland under the US-Ireland R\&D Partnership Programme (USI 175). G. G. acknowledges support from FQXi and DFG FOR2724 and also from the European Union Horizon 2020 research and innovation programme under the Marie Sk\l{}odowska-Curie grant agreement No. 101026667.

\bibliography{draft}

\end{document}